\documentclass[namedreferences]{crckbked}

\usepackage[dvips]{graphicx}
\begin{document}

\setcounter{secnumdepth}{1}
\setcounter{tocdepth}{1}

\typeout{^^J*** Kluwer Academic Publishers - prepress department^^J*** 
    Documentation for book editors using the Kluwer class files}
\begin{article}
\begin{opening}

\title{{\ } \\[1mm]
Constraints on  Non-Newtonian  Gravity
from Recent \protect{\\}
Casimir  Force  Measurements}

\author{V.M.~MOSTEPANENKO}
\institute{Departamento de F\'{\i}sica, Universidade Federal 
da Para\'{\i}ba,\\
C.P.~5008, CEP 58059-970,
Jo\~{a}o Pessoa, Pb-Brazil. \\
On leave from
A.~Friedmann Laboratory for Theoretical Physics,
St.Petersburg, Russia 
}

\begin{abstract}
Corrections to Newton's gravitational law inspired by extra dimensional
physics and by the exchange of light and massless elementary
particles between the atoms of two macrobodies are considered. These
corrections can be described by the potentials of Yukawa-type and
by the power-type potentials with different powers. The strongest
up to date constraints on the corrections to Newton's gravitational law
are reviewed following from the E\"{o}tvos- and Cavendish-type
experiments and from the measurements of the Casimir and
van der Waals force.  We show that the recent measurements
of the Casimir force gave the possibility to strengthen the
previously known constraints on the constants of hypothetical 
interactions up to several thousand times in a wide
interaction range. Further strengthening is expected in near future
that makes Casimir force measurements a prospective test for the
predictions of fundamental physical theories. 
\end{abstract}
\end{opening} 

\section{Introduction}

\noindent
It is common knowledge that  
the gravitational interaction is described on a different basis
than all the other physical interactions. Up to the present
there is no unified description of gravitation and gauge
interactions of the Standard Model which would be satisfactory
both physically and mathematically.
Gravitational interaction persistently avoids unification
with the other interactions. In addition,
there is an evident lack of experimental data in gravitational physics.
Newton's law of
gravitation, which is also valid with high precision in the framework
of the Einstein General Relativity Theory, is not verified
with a sufficient precision
at the separations less than 1\,mm. Surprisingly, at the separations less 
than 1$\,\mu$m corrections to the Newton's gravitational law are not excluded
experimentally that are many orders of magnitude
greater than the Newtonian force itself.
What this means is the general belief, that the Newton's law of gravitation
is obeyed up to Planckean separation distances, is nothing more than a large
scale extrapolation. It is meaningful also that the Newton's gravitational
constant $G$ is determined with much less accuracy than the other
fundamental physical constants. In spite of all attempts the results of
recent experiments on the precision measurement of $G$ are 
contradictory [{1}].

Prediction of non-Newtonian corrections to the law of gravitation comes
from the extra dimensional unification schemes of High Energy Physics.
According to this schemes, which go back to Kaluza [{2}] and
Klein [{3}], the true dimensionality of physical space is larger than 3 
with the extra dimensions being spontaneously compactified at the Planckean
length-scale. At the separation distances several times larger than a 
compactification scale, the Yukawa-type corrections to the Newtonian 
gravitational potential do arise. This prediction would be of only 
academic interest if to take account of the extreme smallness of the
Planckean length $l_{Pl}=\sqrt{G}\sim 10^{-33}\,$cm (we use units
with $\hbar=c=1$) and the excessively high value of the Planckean
energy $M_{Pl}=1/\sqrt{G}=10^{19}\,$GeV.
Recently, however, the low energy (high compactification length)
unification schemes were proposed [{4,5}]. In the  framework of
these schemes the ``true'', multidimensional, Planckean energy
takes a moderate value $M_{\ast}\sim 10^3\,$GeV=1\,TeV and the value 
of a compactification
scale belongs to a submillimeter range. It is amply clear that in the 
same range the Yukawa-type corrections to the Newtonian gravitation
are expected [{6,7}] and this prediction can be verified
experimentally.

Much public attention given to non-Newtonian gravitation is
generated not only by the extra dimensional physics. The new
long-range forces which can be considered as corrections to the
Newton's law of gravitation
are produced also by the exchange of light and
massless hypothetical elementary particles between the atoms of
closely spaced macrobodies. Such particles (like axion, scalar axion,
dilaton, graviphoton, moduli, arion etc.) 
are predicted by many extensions to
the Standard Model and practically inavoidable in the modern
theory of elementary particles and their interactions [{8}].
The long-range forces produced due to the exchange of hypothetical
particles can be considered as some corrections to the Newton's
gravitational law
 leading to the same phenomenological consequences as in the 
case of extra spatial dimensions.

The constraints on the constants characterizing the magnitude and
interaction range of hypothetical forces are usually obtained from
the gravitational experiments of Cavendish- and E\"{o}tvos-type.
These experiments lead to the most strong constraints in the
interaction range
$10^{-5}\,\mbox{m}<\lambda<10^{6}\,$km (see Ref.~[{9}] 
and also some
recent results in Refs.~[{10--13}]). 
In nanometer and micrometer interaction range the best constraints
on the constants of hypothetical interactions follow from the
van der Waals and Casimir force measurements which provide the
dominant background force at so small separations. The first
results in this direction were obtained in Refs.~[{14,15}] 
(see also
Refs.~[{16,17}] for details).

During the last years,
the new experiments on measuring the Casimir force with
an increased precision were performed [{18--25}].
They gave the possibility to considerably increase the strength of
constraints on hypothetical interactions within a submillimeter
interaction range [{26--32}].
Thus, from the measurement of the Casimir force by the use of
an atomic force microscope [{19--21}] the strengthening of the
previously known constraints up to 4500 times was obtained
(the dynamical Casimir force measurements [{23,33}] lead to weaker
constraints than those mentioned above). The increased
experimental precision calls for a more accurate theory taking
into account corrections to the Casimir force due to surface roughness,
finite conductivity of the boundary metal and nonzero temperature.
New constraints were obtained from a comparision between the more
precise experimental data and improved theory (for the recent review
of both experimental and theoretical developments in the Casimir
effect see Ref.~[{34}]).

In the present paper we report the most strong constraints on the
hypothetical long-range interactions obtained from the Casimir
effect. In Sec.~2 the hypothetical long-range forces are discussed
originating from both extra dimensional physics and exchange of light
elementary particles predicted by the unified gauge theories of
fundamental interactions. 
In Sec.~3 the constraints from gravitational experiments are
briefly summarized.
Sec.~4 contains constraints following
from Lamoreaux experiment [{18}] on measuring the Casimir force
by means of a torsion pendulum. In Sec.~5 Mohideen et al
experiments [{19,20}] on measurement of the Casimir force
between an aluminum disk and a sphere by means of an atomic force
microscope are considered. Sec.~6 is devoted to the results of
Ederth experiment [{22}]. In Sec.~7 the most conclusive and
reliable results are presented following from Mohideen et al 
experiment on measuring the Casimir force between gold surfaces
[{21}]. In Sec.~8 reader will find the most recent results
obtained from the lateral Casimir force measurement and from
the new experiment using a microelectromechanical torsional
oscillator.
Sec.~9 contains conclusions and discussion.

Throughout the paper units are used in which $\hbar=c=1$.

\section{Origination of the hypothetical long-range interactions}

\noindent
The usual Newton's law of gravitation is only valid in a
4-dimensional space-time. If the extra dimensions exist, it will
be modified by some corrections. In models with large but compact
extra dimensions (like those proposed in Ref.~[{4}]) the gravitational
potential between two point particles with masses $m_{\>1}$ and $m_2$
separated by a distance $r\gg R_{\ast}$, where $R_{\ast}$ is a
compactification scale, is given by [{6,7}]
\begin{equation}
V(r)=-\frac{Gm_1m_2}{r}\left(1+\alpha_Ge^{-r/\lambda}\right).
\label{e1}
\end{equation}
\noindent
The first term in the right-hand side of Eq.~({1}) is the
Newtonian contribution, whereas the second term represents the Yukawa-type
correction. Here $G$ is the Newton's gravitational constant,
$\alpha_G$ is a dimensionless constant depending on the nature of extra
dimensions and $\lambda$ is the interaction range of a correction.

The dimensionless constant $\alpha_G$ in ({1}) depends on the
nature of the extra dimensions. By way of example, for
a toroidal compactification with all extra dimensions having
equal size, $\alpha_G=2n$ [{6,7}].
If extra dimensions have the topology of $n$-sphere 
$\alpha_G=n+1$ [{6,7}]. 

In fact the search of corrections to Newtonian gravity, like in
Eq.~({1}), is the simplest way to check the predictions of the
models with low compactification scale. The thing is that, 
according to these models, all interactions and particles of 
the Standard
Model are considered as living on a $(3+1)$-dimensional wall.
They remain almost unchanged as this wall has a thickness only of
order $M_{\ast}^{-1}\sim 10^{-17}\,$cm in the extra dimensions.
Only gravitational interaction penetrates freely into extra
dimensions and can serve as a test for their existence.

At small separation distances $r\ll R_{\ast}$ the usual Newton's law
of gravitation should be generalized to
\begin{equation}
V(r)=-\frac{G_{4+n}m_1m_2}{r^{n+1}}
\label{e2}
\end{equation}
\noindent
in order to preserve the continuity of the force lines in a
$(4+n)$-dimensional space-time. Here $G_{4+n}$ is the underlying
multidimensional gravitational constant connected with the usual one
by the relation $G_{4+n}\sim GR_{\ast}^n$.

In fact the characteristic energy scale in multidimensional space-time
is given by the multidimensional Planckean mass 
$M_{\ast}=1/G_{4+n}^{1/(2+n)}$, and the compactification scale
is given by [{4}]
\begin{equation}
R_{\ast}=\frac{1}{M_{\ast}}\left(\frac{M_{Pl}}{M_{\ast}}\right)^{2/n}
\sim 10^{\frac{32}{n}-17}\,\mbox{cm},
\label{e3}
\end{equation}
\noindent
where $M_{Pl}=1/\sqrt{G}$ is the usual Planckean mass, 
$n\geq 1$, and
$M_{\ast}\sim 10^3\,$GeV as was told in Introduction. 
Then, for $n=1$ (one
extra dimension) one finds from Eq.~({3}) 
$R_{\ast}\sim 10^{15}\,$cm.
If to take into account that, as was shown in Refs.~[{6,7}],
$\alpha_G\sim 10$ and $\lambda\sim R_{\ast}$, 
this possibility must be
rejected on the basis of the solar system tests of Newton's 
gravitational law [{9}].
If, however, $n=2$ one obtains from Eq.~({3}) $R_{\ast}\sim 1\,$mm,
and for $n=3$ $R_{\ast}\sim 5\,$nm. For these scales the corrections of 
form ({1}) to Newton's gravitational law 
are not excluded experimentally.

The other type of multidimensional models considers noncompact but warped
extra dimensions. In these models the leading contribution to the
gravitational potential is given by [{5,35}]
\begin{equation}
U(r)=-\frac{Gm_1m_2}{r}\left(1+\frac{2}{3k^2r^2}\right),
\label{e4}
\end{equation}
\noindent
where $r\gg 1/k$ and $1/k$ is the so-called warping scale. Here 
the correction 
to the Newton's gravitational law depends on the separation distance inverse
proportionally to the third power of separation.

As was mentioned in Introduction, many extensions to the Standard Model predict
the hypothetical long-range forces, distinct from gravitation and
electromagnetism, caused by the exchange of light and massless elementary
particles between the atoms of macrobodies. Under appropriate
parametrization of the interaction constant these forces also can be
considered as some corrections to the Newton's gravitational law.
The velocity independent part of the effective potential due to the
exchange of hypothetical particles between two atoms can be calculated
by means of Feynman rules. For the case of massive particles with
mass $\mu=1/\lambda$ ($\lambda$ is their Compton wavelength) the effective
potential takes the Yukawa form
\begin{equation}
V_{Yu}(r)=-\alpha N_1N_2\frac{1}{r}e^{-r/\lambda},
\label{e5}
\end{equation}
\noindent
where $N_{1,2}$ are the numbers of nucleons in the atomic nuclei,
$\alpha$ is a dimensionless interaction constant. If to introduce
a new constant $\alpha_{\>G}=\alpha/(Gm_p^2)\approx 1.7\times 10^{38}\alpha$
($m_p$ being a nucleon mass) and consider the sum of potential ({5})
and Newton's gravitational potential one returns back to the potential
({1}).

For the case of exchange of one massless particle the effective potential
is just the usual Coulomb potential which is inverse proportional to
separation. The effective potentials inverse proportional to higher
powers of a separation distance appear if the exchange of even number
of pseudoscalar particles is considered. The power-type potentials with
higher powers of a separation are obtained also in the exchange of
two neutrinos, two goldstinos or other massless fermions [{16}].
The resulting interaction potential acting between two atoms can
be represented in the form [{36}]
\begin{equation}
U(r)=-\Lambda_lN_1N_2\frac{1}{r}\left(\frac{r_0}{r}\right)^{l-1},
\label{e6}
\end{equation}
\noindent
where $r_0=1\,$F=$10^{-15}\,$m is introduced for the proper
dimensionality of potentials with different $l$, and $\Lambda_{\>l}$
with $l=1,\,2,\,3,\ldots$ are the dimensionless constants.

If to introduce a new set of constants 
$\Lambda_{\>l}^{G}=\Lambda_l/(Gm_p^2)$ and consider the sum of ({6})
and Newton's gravitational potential one obtains
\begin{equation}
U_l(r)=-\frac{Gm_1m_2}{r}\left[1+\Lambda_l^{\! G}
\left(\frac{r_0}{r}\right)^{l-1}\right].
\label{e7}
\end{equation}

\noindent
This equation represents the power-type hypothetical interaction as a
correction to the Newton's gravitational law. The potential ({4})
following from the extra dimensional physics is obtained from 
Eq.~({7}) with $l=3$. Note that the case $l=3$ corresponds also to 
two arions exchange between electrons [{16}].

\section{Constraints from gravitational experiments}

\noindent
Constraints on the corrections to Newton's gravitational law can be obtained
from the experiments of E\"{o}tvos- and Cavendish-type. In the 
E\"{o}tvos-type experiments the difference between inertial and 
gravitational masses of a body is measured, i.e. the equivalence
principle is verified. The existence of an additional hypothetical
force which is not proportional to the masses of interacting bodies can
lead to the appearance of the effective difference between inertial and
gravitational masses. Therefore some constraints on hypothetical
interactions emerge from the experiments of E\"{o}tvos type.

The typical result of the E\"{o}tvos-type experiments is that the relative 
difference between the accelerations imparted by the Earth, Sun or some
laboratory attractor to various substances of the same mass is less
than some small number. Many E\"{o}tvos-type experiments were performed 
(see, e.g., Refs.~[{37--40}]). 
By way of example, in Ref.~[{39}] the above
relative difference of accelerations was to be less than $10^{-11}$.

The results of the two precise E\"{o}tvos-type experiments can be
found in Refs.\ [{10,41}]. They permit to ob\-tain the best 
constraints on the
constants of hypothe\-ti\-cal long-range interactions inspired by extra
dimensions or by the exchange of light and massless elementary
particles (see Fig.~1).

The constraints under consideration can be obtained also from the
Cavendish-type experiments. In these experiments the deviations of the
gravitational force $F$ from Newton's law are measured (see, e.g.,
Refs.~[{42--47}]). 
The characteristic value of deviations in the case of
two point-like bodies a distance $r$ apart can be described by
the parameter
\begin{equation}
\varepsilon=\frac{1}{rF}\frac{d}{dr}\left(r^2F\right),
\label{e8}
\end{equation}
\noindent
which is equal exactly to zero in the case of pure Newton's
gravitational force.
For example, in Refs.~[{44,45}] $\vert \varepsilon\vert\leq 10^{-4}$
at the separation distances $r\sim 10^{-2}-1\,$m. This can be used to
constrain the size of corrections to the Newton's gravitational law.
The results of one of most recent Cavendish-type experiments can be
found in Ref.~[{11}].

\begin{figure}[t]
\vspace*{-8.5cm}
\centerline{
\includegraphics{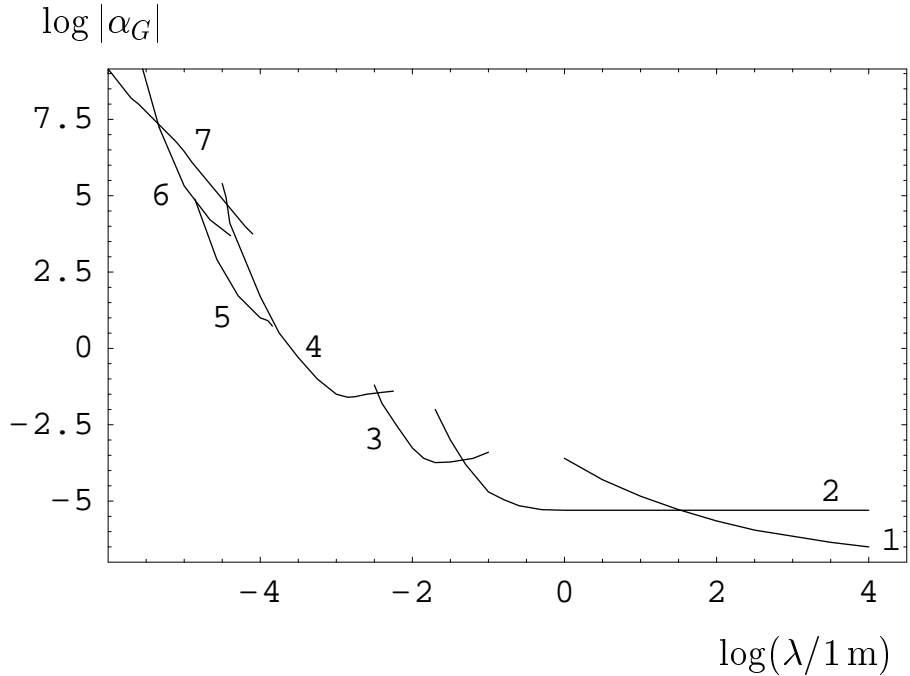}
}
\vspace*{-11.cm}
\caption{Constraints on the Yukawa-type corrections to Newton's
gravitational law. Curves 1,\,2 follow from the E\"{o}tvos-type experiments, 
and curves 3--6 follow from the Cavendish-type experiments.
The beginning of curve 7 shows constraints from the measurements
of the Casimir force. 
Permitted regions on ($\lambda,\,\alpha_G$)-plane
lie beneath the curves.
}
\end{figure}

Let us now outline the strongest constraints on the corrections to
Newton's gravitational law obtained up to date from the gravitational
experiments. The constraints on the parameters of Yukawa-type 
correction, given by Eq.~({1}), are presented in Fig.~1.
In this figure, the regions of $(\lambda,\,\alpha_G)$-plane above
the curves are prohibited by the results of the experiment under
consideration, and the regions below the curves are permitted.
By the curves 1 and 2 the results of the best E\"{o}tvos-type 
experiments are shown (Refs.~[{41}] and 
[{10}], respectively). Curve 4 
represents constraints obtained from the Cavendish-type experiment 
of Ref.~[{11}]. At the intersection of curves 2 and 4 the better
constraints are given by curve 3 following from the results of
older Cavendish-type experiment of Ref.~[{47}]. As is seen from Fig.~1,
rather strong constraints on the Yukawa-type corrections to
Newton's gravitational law ($\alpha_G<10^{-5}$) are obtained only
within the interaction range $\lambda>0.1\,$m.
With decreasing $\lambda$ the strength of constraints falls off, so
that at $\lambda=0.1\,$mm $\alpha_G<100$.
By the beginning of curve 7 the constraints are shown following from
the Casimir force measurements (see Sec.~4).

Recently two more precise Cavendish-type experiments were performed
[12, 13] by the use of the micromachined torsional oscillator.
They have permitted significantly increase the strength of
constraints on $\alpha_{\>G}$ in the interaction range around
(10--100)\,$\mu$m (see curve 5 [{12}] and curve 6 [{13}]).

Now we consider constraints on the power-type corrections to
Newton's gravitational law given by Eq.~({7}). The best of them follow
from the E\"{o}tvos- and Cavendish-type experiments. They are collected
in Table 1.
\begin{table}[h]
\caption{Constraints on the constants of power-type potentials.}
{\begin{tabular}{clll} %
\hline
$l$ & $\vert\Lambda_l\vert_{\max}$ & $\vert\Lambda_l^{\! G}\vert_{\max}$ &
Source \\%
\hline
1 & $6\times 10^{-48}$& $1\times 10^{-9}$ & Ref.~[48] \\
2 & $2.4\times 10^{-30}$& $4\times10^8$& Ref.~[10] \\
3& $7\times 10^{-17}$& $1.2\times 10^{22}$& Refs.~[47,\,49,\,50] \\
4& $7.5\times 10^{-4}$& $1.3\times 10^{35}$& Refs.~[45,\,50]\\ 
5&$1.2\times 19^9$&$2\times 10^{47}$ &
Refs.~[45,\,50]\\
\hline
\end{tabular}}
\end{table}

For $l=1,\,2$ the constraints presented in Table 1 are obtained
from the E\"{o}tvos-type experiments, and for $l=3,\,4,\,5$ from the
Cavendish-type ones. It is seen that the strength of constraints falls 
greatly with the increase of $l$.

\section{Constraints following from Lamoreaux experiment
by means of torsion pendulum}

\noindent
As is seen from Sec.~3, for larger interaction distances the best
constraints on the corrections to Newton's gravitational law follow
from the E\"{o}tvos-type experiments and for lesser interaction 
distances from
the Cavendish-type ones. With the further decrease of the 
characteristic interaction distance
the strength of constraints following from the gravitational experiments
greatly reduces. Within a micrometer separations, the Casimir and van
der Waals force [{17,51,52}] 
becomes the dominant force between two
macrobodies. As was shown in Ref.~[{14}] for the case of Yukawa-type
interactions with a micrometer interaction range 
and in Ref.~[{15}] for the power-type ones, the measurements of
the van der Waals and Casimir forces lead to the strongest constraints on
non-Newtonian gravity
(see the discussion about the Casimir effect as a test for
non-Newtonian gravitation in Ref.~[{53}]).

Currently a lot of precision experiments on the measurement of the
Casimir and van der Waals force has been performed 
(see Ref.~[{34}] for a review).
As was mentioned in Introduction,
the extensive theoretical study of different corrections to the Casimir
force due to surface roughness, finite conductivity of a boundary metal
and nonzero temperature gave the possibility to compute  the theoretical
value of this force with high precision. At the moment the agreement
between theory and experiment at a level of 1\% is achieved for the
smallest experimental separation distances [{34}]. This permitted
to obtain stronger constraints on the corrections to Newton's 
gravitational law
from the results of the Casimir force 
measurements [{26--32,54,55}].
Here we briefly present the strongest constraints of this type
starting from the first modern experiment performed by
Lamoreaux [{18}].

In Ref.~[{18}] the Casimir force between a spherical lens and a disk
made of quartz (with the densities
$\rho^{\>\prime}=2.23\times 10^3\,{\mbox{kg/m}}^3$ and
$\rho =2.4\times 10^3\,{\mbox{kg/m}}^3$, respectively)
and coated by $Cu$ and $Au$ layers of thickness
$\Delta_{\>1}=\Delta_2=0.5\,\mu$m (with the densities
$\rho_{1}=8.96\times 10^3\,{\mbox{kg/m}}^3$,
$\rho_2=19.32\times 10^3\,{\mbox{kg/m}}^3$)
was measured by the use of torsion pendulum. 
The disk radius was
$L=1.27\,$cm and a lens height and curvature radius were
$ H=0.18\,$cm and $R=12.5\,$cm, respectively.

The absolute error of force measurements in Ref.~[18] was about
$\Delta F=10^{-11}\,$N for the separation range between a disk
and a lens 
$1\,\mu\mbox{m}\leq a\leq 6\,\mu$m.
In the limits of this error the theoretical expression
for the Casimir force was confirmed
\begin{equation}
F^{(0)}(a)=-\frac{\pi^3}{360}\,\frac{R}{a^3}.
\label{e9}
\end{equation}

No corrections to Eq.~({9}) due to surface roughness,
finite conductivity of the boundary metal or nonzero
temperature were reported. These corrections, however, may
not lie within the limits of the absolute error $\Delta F$.
By way of example,
 at $a=1\,\mu$m roughness correction $\Delta_RF(a)$
may be around 12\% 
of $F^{(0)}$ or even larger [{26}]. The
finite conductivity correction $\Delta_{\delta_0}F(a)$
for gold surfaces at $a=1\,\mu$m separation is
10\% of $F^{(0)}$ [{56}]. (Note that $\Delta F$ is
around 3\% of $F^{(0)}$  at $a=1\,\mu$m.)
As to temperature correction $\Delta_T F(a)$,
it achieves 174\% of $F^{(0)}$ at the separation $a=6\,\mu$m,
where, however, $\Delta F$  is around 700\% 
of $F^{(0)}$.
For this reason, the constraints on the Yukawa-type interaction
following from Lamoreaux experiment were found from the
inequality [{26}]
\begin{equation}
|F_{th}(a)-F^{(0)}(a)|\leq\Delta F,
\label{e10}
\end{equation}
\noindent
where $F_{th}$ is the theoretical force value including $F^{(0)}$,
all the corrections to it mentioned above, and also the
hypothetical Yukawa-type interaction
\begin{equation}
F_{th}(a)=F^{(0)}(a)+\Delta_R F(a)+\Delta_{\delta_0} F(a)
+\Delta_T F(a)+F_{Yu}(a)
\label{e11}
\end{equation}
\noindent
(we remind that the sign of a finite conductivity correction is
opposite to the sign of other corrections).

The hypothetical interaction in a configuration of a spherical
lens above a disk was computed in Ref.~[{26}]. 
For $\lambda$ smaller
or of order of separation $a$ the result is given by
\begin{eqnarray}
&&
F_{Yu}(a)=-4\pi^2\alpha_G G\lambda^3 e^{-a/\lambda}R
\label{e12} \\
&&\phantom{aaa}
\times
\left[\rho_2
-\left(\rho_2-\rho_1\right)
e^{-\Delta_2/\lambda}
-\left(\rho_1-\rho^{\prime}\right)
e^{-(\Delta_2+\Delta_1)/\lambda}\right]
\nonumber \\
&&\phantom{aaa}
\times
\left[\rho_2-\left(\rho_2-\rho_1\right)
e^{-\Delta_2/\lambda}-\left(\rho_1-\rho\right)
e^{-(\Delta_2+\Delta_1)/\lambda}\right].
\nonumber
\end{eqnarray}
\noindent
For larger $\lambda$, $F_{Yu}(a)$ was
computed numerically [{26}].

The obtained constraints [{26}] are shown in Fig.~2 (curve 7,a for 
$\alpha_G>0$ and curve 7,b for $\alpha_G<0$). In this figure, the regions
of $(\alpha_G,\lambda)$-plane above the curves are prohibited, and
the regions below the curves are permitted by the results of an
experiment under consideration. By the curves 6  the results
of the best Cavendish-type experiments are shown 
(Ref.~[{13}]). Curve 8 represents constraints obtained from the
Casimir force measurements between dielectrics [{16,17}].
Line 13 demonstrates the typical prediction of extra dimensional
theories. 
The strengthening of constraints given by curves 7,a and 7,b 
comparing
curve 8 is up to a factor of 30 
in the interaction range
$2.2\times 10^{-7}\,\mbox{m}\leq\lambda\leq 5\times 10^{-6}\,$m
(a weaker result was obtained later in 
Ref.~[{29}] where the corrections to
the ideal Casimir force of Eq.~({9}) were not taken into account). 
This shows that the Casimir force measurements are competitive with
the Cavendish-type experiments in a micrometer interaction range.

\begin{figure}[t]
\vspace*{-8.5cm}
\centerline{
\includegraphics{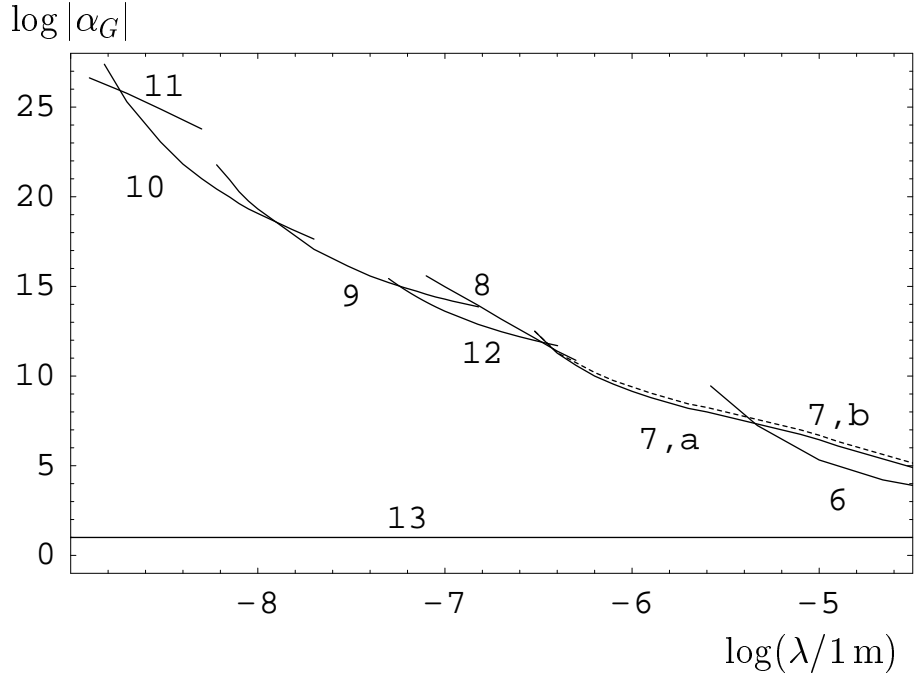}
}
\vspace*{-10.5cm}
\caption{Constraints on the Yukawa-type corrections to Newton's
gravitational law. Curves 8--10,\,12 follow from the Casimir, 
and curve 11
from the van der Waals force measurements.
The typical prediction of extra dimensional physics is shown 
by curve 13.
}
\end{figure}

\section{Constraints following from Mohideen et el experiments
with $Al$ surfaces by means of atomic force microscope}

\noindent
A major progress in obtaining more strong constraints on the
Yukawa-type interactions within a nanometer range was achieved
due to the measurements of the Casimir force by means of an atomic force
microscope [{19--21}]. 
In Refs.~[{19,56}] the results of the Casimir
force measurement between a flat supphire disk
($L=0.625\,$cm,
$\rho=4.0\times 10^3\,{\mbox{kg/m}}^3$)
and a polystyrene sphere ($R=98\,\mu$m,
$\rho^{\>\prime}=1.06\times 10^3\,{\mbox{kg/m}}^3$) were presented
in comparison with a complete theory taking into account the finite
conductivity and roughness corrections. Temperature corrections are
not essential in the separation range
$0.12\,\mu\mbox{m}\leq a\leq 0.9\,\mu$m used in Refs.~[{19,56}].
The test bodies were coated by the aluminum layer
($\rho_{\>1}=2.7\times 10^3\,{\mbox{kg/m}}^3$) of
$\Delta_1=300\,$nm thickness and $Au/Pd$ layer
($\rho_{\>2}=16.2\times 10^3\,{\mbox{kg/m}}^3$) of the thickness
$\Delta_2=20\,$nm (this latter was used to prevent the oxidation
processes; it is almost transparent for electromagnetic oscillations
of characteristic frequency). The absolute error of force measurements 
in Refs.~[{19,56}] was
$\Delta F=2\times 10^{-12}\,$N.
In the limits of this error the theoretical expression for the Casimir
force with corrections to it due to the surface roughness and finite
conductivity was confirmed.

In the improved version of this experiment [{20}] the $Au/Pd$ layer
was made thiner ($\Delta_2=7.9\,$nm) and the other experimental
parameters were as follows: $\Delta_1=250\,$nm, $L=0.5\,$cm,
$R=100.85\,\mu$m, $100\,\mbox{nm}\leq a\leq 500\,$nm.
Due to experimental improvements like the use of vibration isolation,
lower systematic errors, independent measurement of a surface separation
and smoother metal surface, the smaller absolute error of force measurement
$\Delta F=1.3\times 10^{-12}\,$N was achieved. In the limits of this
error experimental data were in agreement with a complete theory.

To obtain constraints on hypothetical interactions, the hypothetical
force was computed [{27,28}] with account of surface roughness
contribution which is especially important at the closest separations
\begin{equation}
F_{Yu}(a)=
\sum\limits_{i}w_iF_{Yu}(a_i).
\label{e13}
\end{equation}
\noindent
Here $w_i$ are the probabilities for different values of a separation
distance between the distorted surfaces, and $F_{Yu}$ is given by
Eq.~({12}). The values of $w_i$ were found [{56}] on the basis of
atomic force microscope measurements of surface roughness. The typical
roughness heights were 40\,nm and 20\,nm (Refs.~[{19,56}]) and
14\,nm and 7\,nm (Ref.~[{20}]).

The constraints were obtained from the inequality
\begin{equation}
|F_{Yu}(a)|\leq\Delta F,
\label{e14}
\end{equation}
\noindent
because the theoretical expression for the Casimir force with all
corrections to it was confirmed experimentally unlike the case
considered in Sec.~4.

Using the above experimental parameters of 
Refs.~[{19,56}], the strengthening
of constraints up to 140 times was obtained [{27}] as compared with
the measurements of the van der Waals and Casimir force between dielectrics.
The strengthening holds within the interaction range
$5.9\,\mbox{nm}\leq\lambda\leq 100\,$nm.

Even stronger constraints were obtained [{28}] from the experiment
of Ref.\ [{20}] in a wider interaction range
$5.9\,\mbox{nm}\leq\lambda\leq 115\,$nm.
The new constraints  are up to 560 times stronger than 
the old ones given by curves 8 and 11 in Fig.~2 which follow
from old mesurements of the Casimir and van der Waals force
between dielectrics (the final constraint curve from the atomic
force microscopy measurements will be obtained in Sec.~7).

The above constraints obtained on the basis of 
Refs.~[19,\,20,\,56] are
found for the closest separation distance $a$ (120\,nm in 
Refs.~[19,\,56] and 100\,nm in Ref.~[20]). If to decrease the minimal
value of $a$, stronger constraints can be obtained.
This is true also if the heavier metal coating is used (see below).

\section{Constraints on hypothetical interactions following
from Ederth \protect{\\} experiment with two crossed cylinders}

\noindent
In Ref.~[{22}] the Casimir force acting between two crossed quartz
cylinders of 1\,cm radius was measured (quartz density
$\rho=\rho^{\>\prime}=2.23\times 10^3\,{\mbox{kg/m}}^3$).
Each cylinder was coated by a layer of $Au$ (density
$\rho_{\>1}=18.88\times 10^3\,{\mbox{kg/m}}^3$ and thickness
$\Delta_1=200\,$nm) and outer layer of hydrocarbon (density
$\rho_{\>2}=0.85\times 10^3\,{\mbox{kg/m}}^3$, thickness
$\Delta_2=2.1\,$nm).
The absolute error of force measurements was 
$\Delta F=10\,$nN which is much larger than in the experiments
discussed above. Within the limits of this error the theoretical
expression for the Casimir force between cylinders was confirmed.
The separation range between the cylinders was in the limits
$20\,\mbox{nm}\leq a\leq 100\,$nm,
i.e. the more close separations were achieved. The other
experimental improvement of Ref.~[{22}] lies in the use of smoother
surfaces. The root mean square roughness of the cylindrical
surfaces was decreased up to 0.4\,nm.

There were also some disadvantages in the experiment of Ref.~[{22}]
as compared with the previous experiments. One of them is connected 
with the presence of hydrocarbon coating which complicates the
independent measurement of the residual electrostatic force.
The other disadvantage is a substantial deformation of the $Au$
coating caused by the attractive forces in contact and by
relatively soft glue used to support the $Au$ layer. As a result,
there is no independent and exact determination of surface
separation in Ref.~[22].

The constraints on the parameters of Yukawa-type interaction,
following from the experiment of Ref.~[{22}], were obtained from
Eq.~({14}) in Ref.~[{30}]. It was shown [{30}] that the
hypothetical force between two cylinders, crossed at a right angle, is
given once more by Eq.~({12}). Surface roughness contribution
is not essential here as the roughness amplitude was considerably
decreased.

The obtained constraints [{30}] are shown by curve 10 in Fig.~2
(by curve 11 the constraints following from the van der Waals
force measurements between dielectrics are demonstrated).
They are up to 300 times stronger than the previously known ones
within the separation range
$1.5\,\mbox{nm}\leq\lambda\leq 11\,$nm. 
This result was obtained at the closest separation
distance $a=20\,$nm.

\section{Constraints on hypothetical interactions following
from Mohideen et el experiments with gold  surfaces}

\noindent
The most conclusive measurement of the Casimir force by means
of atomic force microscope was performed between a sapphire disk
and polystyrene sphere ($R=95.65\,\mu$m) coated by $Au$ layer
of $\Delta_1=86.6\,$nm thickness [{21}]. No additional coating
was used which added complexity to interpretation of experimental
data of Refs.~[{19,20,22}]. Some other improvements were
implemented also in this experiment. Specifically, the root
mean square amplitude of surface roughness was decreased up to
$1.0\pm 0.1\,$nm which is comparable with 
Ref.~[{22}] (see the preceding
section) but did not require additional hydrocarbon coating.
The electrostatic forces were reduced to a value much smaller of
the Casimir force at the shortest separation and used for an
independent measurement of surface separation. Also the measurement
was performed over smaller separations
$62\,\mbox{nm}\leq a\leq 350\,$nm than in previous measurements
by means of atomic force microscope.

The absolute error of force measurements in Ref.~[{21}],
$\Delta F=3.5\times 10^{-12}\,$N, was a bit larger than that in
Refs.~[{19,20}]. 
This was caused by the poor thermal conductivity of the
cantilever resulting from the thiner metal coating used. The increase
of $\Delta F$ is, however, compensated for by the greater increase
of the Casimir force at smaller separations.

Constraints on the constants of Yukawa-type interaction were 
obtained [{31,57}] from Eq.~({14}) using the agreement of
experimental data with a theoretical Casimir force.
Hypothetical force was computed by Eq.~({12}) having regard to
$\Delta_{\>2}=\rho_2=0$. The computational results are shown by curve 9
in Fig.~2. The obtained constraints are stronger up to 19 times,
comparing the previous experiments using the atomic force microscope,
within the interaction range
$4.3\,\mbox{nm}\leq\lambda\leq 150\,$nm.
If to compare with the experiment of 
Ref.~[{22}], the constraints following
from Mohideen et al experiment with $Au$ surfaces prove to be the best
ones in the interaction range
$11\,\mbox{nm}\leq\lambda\leq 150\,$nm.
As a consequence, the constraints, which are up to 4500 times more
stringent than those from older Casimir and van der Waals force
measurements between dielectrics, are obtained from the experiments
by means of the atomic force microscope.

\section{Constraints from measurements of the lateral Casimir
force and from experiment using a microelectromechanical
torsional oscillator}

\noindent
In 2002, the new physical phenomenon, the lateral Casimir force,
was demonstrated first [{24,25}] acting be\-t\-ween a sinusoidally
corrugated gold plate and large sphere. This force acts in a direction
tangential to the corrugated surface. The experimental setup was
based on the atomic force microscope specially adapted for the
measurement of the lateral Casimir force. The measured force oscillates
sinusoidally as a function of the phase difference between the two
corrugations in agreement with theory with an amplitude of
$3.2\times 10^{-13}\,$N at a separation distance 221\,nm.
So small value of force amplitude measured with a resulting absolute error
$0.77\times 10^{-13}\,$N [{25}] with a 95\% confident probability
gives the opportunity to obtain constraints on the respective lateral
hypothetical force which may act between corrugated surfaces.

The obtained constraints [{25,32}] are shown in Fig.~3 as the solid curve.
In the same figure, the short-da\-shed curve indicates constraints 
obtained from the old Casimir force measurements between dielectrics
(curve 8 in Fig.~2), and the long-dashed curve follows from the most
precision measurement of the normal Casimir force between gold surfaces
[{21}] (these constraints were already shown by curve 9 in Fig.~2).
The constraints obtained by means of the lateral Casimir force measurement
are of almost the same strength as the ones known previously in the
interaction range 80\,nm$<\lambda <$150\,nm. However, with the increase of 
accuracy of the lateral Casimir force measurements more promising
constraints are expected.

\begin{figure}[t]
\vspace*{-6.5cm}
\hspace*{-2.5cm}\includegraphics{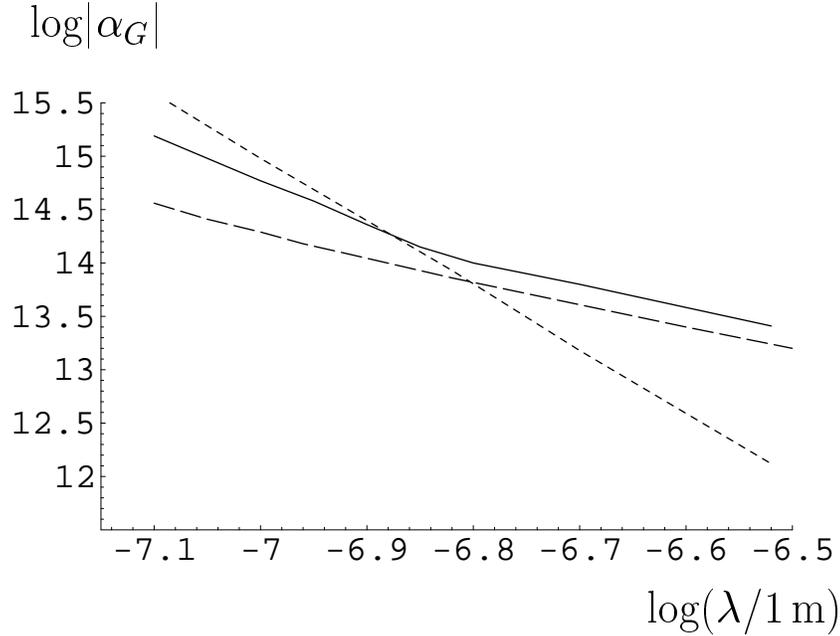}
\vspace*{-11.5cm}
\caption{Constraints on the Yukawa-type corrections to Newton's
gravitational law from the measurement of the lateral Casimir force
between corrugated surfaces (solid curve).
For comparison the short-dashed and long-dashed curves reproduce
curves 8 and 9 of Fig.~2, respectively, obtained from the meausrements
of the normal Casimir force between dielectrics and between gold
surfaces.
}
\end{figure}

Recently one more experiment was performed on measuring the normal
Casi\-mir force. For this purpose the microelectromechanical torsional
oscillator has been used. This permitted to perform measurements
of the Casimir force between a sphere and a plate with   
an absolute error of 0.3\,pN and of the Casimir pressure between
two parallel plates with an absolute error of about 0.6\,mPa
[{58,59}] for separations $0.2-1.2\,\mu$m.
As a result, the new constraints on the Yukawa-type corrections
to Newton's law of gravitation were obtained [{59}] which
are more than one order of magnitude stronger than the
previously known ones within a wide interaction range from
56\,nm to 330\,nm. These constraints are shown by curve 12 in
Fig.~2. It is notable that the constraints given by curve 12
almost completely cover the gap between the modern constraints
obtained by means of an atomic force microscope and a torsion
pendulum. Within this gap the constraints found from the old
measurements of the Casimir force between dielectrics (curve 8 in
Fig.~2) were the best ones. Now they are changed to the more
precise and reliable constraints obtained from the Casimir
force measurements between metals by means of a
microelectromechanical torsional oscillator. 

As is seen from Figs.~2,\,3, the present strength of constraints 
is not 
sufficient to confirm or to reject the predictions of extra dimensional 
physics with the compactification scale $R_{\ast}<0.1\,$mm
(line 13 in Fig.~2).
However, Fig.~2 gives the possibility to set constraints on the parameters
of light hypothetical particles, moduli, for instance. Such particles
are predicted in superstring theories and are characterized by the
interaction range from one micrometer to one centimeter [{60}].

\section{Conclusions and discussion}
\noindent
In the above, the modern constraints on the constants of hypothetical
interactions are reviewed following from the Casimir force
measurements and gravitational experiments.
As is evident from Fig.~2, for separations smaller than $10^{-4}\,$m
much work is needed to achieve the strength of Yukawa interaction
predicted by extra dimensional physics ($\alpha_G\sim 10$).
However, one should remember that Casimir force experiments are also 
sensitive to other non-extra dimensional effects such as exchange
by light elementary particles which can lead to the Yukawa-type forces
with $\alpha_G\gg 10$ (see Refs.~[{9,16,60}]). Therefore, all
experiments which can strengthen the constraints on constants of the
Yukawa-type interaction are of immediate interest to both elementary
particle physics and gravitation [{61}].

The following conclusions might be formulated.
The idea of hypothetical interactions has gained recognition.
New long-range forces additional to the usual gravitation and
electromagnetism are predicted by the extra dimensional physics.
They may be caused also by the exchange of light elementary
particles predicted by the unified gauge theories of fundamental
interactions.

The modern measurements of the Casimir force already gave the
possibility to strengthen constraints on hypothetical
long-range forces up to several thousand times in a wide
interaction range from 1 nanometer to 100 micrometers.

Further strengthening of constraints on non-Newtonian gravity
and other hypothetical long-range interactions from the Casimir
effect is expected in the future. In this way the Casimir force
measurements are quite competitive with the modern accelerator 
and gravitational experiments as a test for predictions of 
fundamental physical theories.

In near future we may expect to obtain the resolution of the problem
are there exist large extra dimensions and the Yukawa-type corrections
to Newtonian gravity at small distances.

\section*{Acknowledgements}
\noindent
The author is grateful  M.~Bordag, R.~Decca, E.~Fischbach, B.~Geyer, 
G.\ L.\ Klimchitskaya, D.\ E.\ Krause, D.~L\'{o}pez, U.\ Mohideen
and M.~Novello for helpful discussions and collaboration.
He thanks V.~de Sabbata and the staff of the ``Ettore
Majorana'' Center for Scientific Culture at Erice for kind
hospitality. 
The partial financial support from CNPq (Brazil) is 
also acknowledged.

\section{References}
\begin{enumerate}
\renewcommand{\theenumi}{\arabic{enumi}.}
\item
Gillies, G.T. (1997) The Newtonian gravitational constant: recent
measurements and related studies,
{\it Rep. Prog. Phys.} {\bf 60}, 151--225.
\item
Kaluza, Th. (1921) On the problem of unity in physics,
{\it Sitzungsber. Preuss. Akad. Wiss. Berlin
Math. Phys.} {\bf K1}, 966.
\item
Klein, O. (1926) Quantum theory and 5-dimensional theory 
of relativity,
{\it Z. Phys.} {\bf 37}, 895.
\item
Arkani-Hamed, N., Dimopoulos, S., and Dvali, G. (1999)
Phenomenology, astrophysics, and cosmology of theories with
submillimeter dimensions and TeV scale quantum gravity, 
{\it Phys. Rev. } {\bf D59},
086004-1--4.
\item
Randall, L. and Sundrum, R. (1999)
Large mass hierarchy from a small extra dimension, 
{\it Phys. Rev. Lett.} {\bf 83}, 3370--3373.
\item
Floratos, E.G. and Leontaris, G.K. (1999)
Low scale unification, Newton's law and extra dimensions,
{\it Phys. Lett. } {\bf B465}, 95-100.
\item
Kehagias, A. and Sfetsos, K. (2000)
Deviations from the $1/r^2$ Newton law due to extra dimensions, 
{\it Phys. Lett. } {\bf B472}, 39--44.
\item
Kim, J. (1987)
Light pseudoscalars, particle physics and cosmology,
{\it Phys. Rep.} {\bf 150}, 1--177.
\item
Fischbach, E. and Talmadge, C.L. (1999)
{\it The Search for Non-Newtonian Gravity}, 
Springer-Verlag, New York.
\item
Smith, G.L., Hoyle, C.D., Gundlach, J.H., Adelberger, E.G.,
Heckel, B.R., and Swanson, H.E. (2000)
Short range tests of the equivalence principle, 
{\it Phys. Rev.} {\bf D61}, 022001-1--20.
\item
Hoyle, C.D., Schmidt, U., Heckel, B.R., Adelberger, E.G.,
Gundlach, J.H.,
Kapner, D.J.,  and Swanson, H.E. (2001)
Submillimeter test of the gravitational inverse-square law: a search
for ``large'' extra dimensions,
{\it Phys. Rev. Lett.} {\bf 86}, 1418--1421.
\item
Long, J.C., Chan, H.W., Churnside, A.B., Gulbis, E.A.,
Varney, M.C.M., and Price, J.C. (2003) Upper limits
to submillimeter range forces from extra space-time dimensions,
{\it Nature} {\bf 421}, 922--925.
\item
Chiaverini, J., Smullin, S.J., Geraci, A.A., Weld, D.M.,
and Kapitulnik, A. (2003) New experimental constraints
on non-Newtonian forces below 100\,$\mu$m,
{\it Phys. Rev. Lett.} {\bf 90}, 151101-1--4.
\item
Kuz'min, V.A., Tkachev, I.I., and Shaposhnikov, M.E. (1982)
Restrictions imposed on light scalar particles by measurements
of van der Waals forces,
{\it JETP Lett. (USA)} {\bf 36}, 59--62.
\item
Mostepanenko, V.M. and Sokolov, I.Yu. (1987)
The Casimir effect leads to new restrictions on long-range
forces constants, 
{\it Phys. Lett. } {\bf A125}, 405--408.
\item
Mostepanenko, V.M. and Sokolov, I.Yu. (1993)
Hypothetical long-range interactions and restrictions on their
parameters from force measurements,
{\it Phys. Rev. } {\bf D47}, 2882--2891.
\item
Mostepanenko, V.M. and Trunov, N.N. (1997) 
{\it The Casimir Effect
and Its Applications}, Clarendon Press, Oxford.
\item
Lamoreaux, S.K. (1997)
Demonstration of the Casimir force in the 0.6 to 6\,$\mu$m
range,
{\it Phys. Rev. Lett.} {\bf 78}, 5--8; (1998) Erratum,
{\bf 81}, 5475.
\item
Mohideen, U. and Roy, A. (1998)
Precision measurement of the Casimir force from 0.1 to
0.9\,$\mu$m, 
{\it Phys. Rev. Lett.} {\bf 81},
4549--4552.
\item
Roy, A., Lin, C.Y., and Mohideen, U. (1999)
Improved precision measurement of the Casimir force,
{\it Phys. Rev.} {\bf D60}, 111101-1--5.
\item
Harris, B.W., Chen, F., and Mohideen, U. (2000)
Precision measurement of the Casimir force using gold surfaces, 
{\it Phys. Rev.} {\bf A62}, 052109-1--5.
\item
Ederth, T. (2000)
Template-stripped gold surface with 0.4-nm rms roughness suitable
for force measurements: Application to the Casimir force in the
20--100\,nm range, 
{\it Phys. Rev. } {\bf A62}, 062104-1--8.
\item
Bressi, G., Carugno, G., Onofrio, R., and Ruoso, G.
(2002) Measurement of the Casimir force between parallel
metallic surfaces, {\it Phys. Rev. Lett.} {\bf 88},
041804-1--4. 
\item
Chen, F., Mohideen, U., Klimchitskaya, G.L., 
and Mos\-te\-panenko, V.M. (2002)
Demonstration of the lateral Casimir force,  
{\it Phys. Rev. Lett.} {\bf 88}, 101801-1--4.
\item
Chen, F.,  Klimchitskaya, G.L., Mohideen, U.,
and Mos\-te\-panenko, V.M. (2002) 
Experimental and theoretical investigation of the lateral
Casimir force between corrugated surfaces,
{\it Phys. Rev. } {\bf A66}, 032113-1--11.
\item
Bordag, M., Geyer, B., Klimchitskaya, G.L., 
and Mos\-te\-panenko, V.M. (1998)
Constraints for hypothetical interactions from a recent
demonstration of the Casimir force and some possible
improvements, 
{\it Phys. Rev. }  {\bf D58}, 075003-1--16.
\item
Bordag, M., Geyer, B., Klimchitskaya, G.L., 
and Mos\-te\-panenko, V.M. (1999)
Stronger constraints for nanometer scale Yukawa-type
hypothetical interactions from the new measurement
of the Casimir force,
{\it Phys. Rev. }  {\bf D60}, 055004-1--7.
\item
Bordag, M., Geyer, B., Klimchitskaya, G.L., 
and Mos\-te\-panenko, V.M. (2000)
New constraints for non-Newtonian gravity in nanometer range 
from the improved precision measurement of the Casimir force,
{\it Phys. Rev. }  {\bf D62}, 011701-1--5.
\item
Long, J.C., Chan, H.W., and Price, J.C. (1999) 
Experimental status of gravi\-ta\-tional-strength forces in
the sub-centimeter range,
{\it Nucl. Phys. } {\bf B539}, 23--34.
\item
Mostepanenko, V.M. and Novello, M. (2001)
Constraints on non-Newto\-nian gravity from the Casimir force
measurement between two crossed cy\-lin\-ders,
{\it Phys. Rev. } {\bf D63}, 115003-1--5.
\item
Fischbach, E., Krause, D.E.,
Mostepanenko, V.M., and Novello, M. (2001)
New constraints on ultrashort-ranged Yukawa interactions
from atomic force microscopy,
{\it Phys. Rev. } {\bf D64}, 075010-1--7.
\item
Klimchitskaya, G.L. and Mohideen, U. (2002)
Constraints on Yukawa-type hypothetical interactions from
recent Casimir force measurements,
{\it Int. J. Mod. Phys. } {\bf A17},
4143--4152.
\item
Carugno, G., Fontana, Z., Onofrio, R., and Ruoso, G. (1997)
Limits on the existence of scalar interactions in the
submillimeter range,
{\it Phys. Rev.} {\bf D55}, 6591--6595.
\item
Bordag, M., Mohideen, U., and Mostepanenko, V.M. (2001)
New developments in the Casimir effect,
{\it Phys. Rep.}  {\bf 353}, 1--205.
\item
Randall, L., and Sundrum, R. (1999) 
An alternative to compactification,
{\it Phys. Rev. Lett.} {\bf 83}, 4690--4693.
\item
Feinberg, G.and Sucher, J. (1979)
Is there a strong van der Waals force between hadrons,
{\it Phys. Rev. } {\bf D20}, 
1717--1735.
\item
Stubbs, C.W., Adelberger, E.G., Raab, F.J., Gundlach, J.H.,
Heckel, B.R., McMurry, K.D., Swanson, H.E., and Watanabe, R. (1987)
Search for an inter\-me\-dia\-te-range interactions,
{\it Phys. Rev. Lett.} {\bf 58}, 1070--1073.
\item
Stubbs, C.W., Adelberger, E.G.,
Heckel, B.R., Rogers, W.F., Swanson, H.E., Watanabe, R.,
Gundlach, J.H., and Raab, F.J. (1989)
Limits on composition-dependent interactions using a laboratory
source --- is there a 5th force coupled to isospin,
{\it Phys. Rev. Lett.} {\bf 62}, 609--612.
\item
Heckel, B.R., Adelberger, E.G., Stubbs, C.W., Su, Y.,
Swanson, H.E., and Smith, G. (1989)
Experimental bounds of interactions mediated by ultralow-mass bosons,
{\it Phys. Rev. Lett.} {\bf 63}, 2705--2708.
\item
Braginskii, V.B. and Panov, V.I. (1972)
Verification of equivalence of inertial and gravitational mass,
{\it Sov. Phys. JETP} {\bf 34}, 463.
\item
Su, Y., Heckel, B.R., Adelberger, E.G., Gundlach, J.H.,
Harris, M., Smith, G.L., and Swanson, H.E. (1994)
New tests of the universality of free fall,
{\it Phys. Rev. } {\bf D50}, 3614--3636.
\item
Holding, S.C., Stacey, F.D., and Tuck, G.J. (1986)
Gravity in mines --- an investigation of Newtonian law,
{\it Phys. Rev. } {\bf D33}, 3487--3497.
\item
Stacey, F.D., Tuck, G.J., Moore, G.I., Holding, S.C.,
Goodwin, B.D., and Zhou, R. (1987)
Geophysics and  the law of gravity,
{\it Rev. Mod. Phys.} {\bf 59}, 157--174.
\item
Chen, Y.T., Cook, A.H., and Metherell, A.J.F. (1984)
An experimental test of the inverse square law of gravitation
at range of 0.1\,m,
{\it Proc. R. Soc. London } {\bf A394}, 47--68.
\item
Mitrofanov, V.P. and Ponomareva, O.I. (1988)
Experimental check of law of gravitation at small distances,
{\it Sov. Phys. JETP} {\bf 67}, 1963.
\item
M\"{u}ller, G., Zurn, W., Linder, K., and Rosch, N. (1989)
Determination of the gravitational constant by an experiment
at a pumped-storage reservoir,
{\it Phys. Rev. Lett.} {\bf 63}, 2621--2624.
\item
Hoskins, J.K., Newman, R.D.,  Spero, R., and Schultz, J. (1985)
Experimental tests of the gravitational inverse-square law
for mass separated from 2 to 105\,cm, 
{\it Phys. Rev. } {\bf D32}, 3084--3095.
\item
Gundlach, J.H., Smith, G.L., Adelberger,  E.G.,
Heckel, B.R., and Swanson, H.E. (1997)
Short-range test of the equivalence principle,
{\it Phys. Rev. Lett.} {\bf 78}, 2523--2526.
\item
Mostepanenko, V.M. and Sokolov, I.Yu (1990)
Stronger restrictions on the constants of long-range forces
decreasing as $r^{-n}$,
{\it Phys. Lett. } {\bf A146}, 373--374.
\item
Fischbach, E. and Krause, D.E. (1999)
Constraints on light pseudoscalars implied by tests of
the gravitational inverse-square law,
{\it Phys. Rev. Lett.} {\bf 83}, 3593--3596.
\item
Milonni, P.W. (1994)
{\it The Quantum Vacuum}, Academic Press,
San Diego.
\item
Milton, K.A. (2001)
{\it The Casimir Effect},
World Scientific, Singapore.
\item
Krause, D.E. and Fischbach, E. (2001)
Searching for extra dimensions and new string-inspired forces in 
the Casimir regime, in C.~L\"{a}mmerzahl, C.W.F.\ Everitt, and
F.W.~Hehl (eds.),
{\it Gyros, Clocks, and
Interferometers: Testing Relativistic Gravity in Space}, 
Springer-Verlag, Berlin, pp.~292--309.
\item
Mostepanenko, V.M. (2002)
Constraints on forces inspired by extra dimensional physics 
following from the Casimir effect,
{\it Int. J. Mod. Phys. } {\bf A17}, 722--731.
\item
Mostepanenko, V.M. (2002)
Experimenatl status of corrections to Newtonian gravitation
inspired by extra dimensions,
{\it Int. J. Mod. Phys. } {\bf A17}, 4307--4316.
\item
Klimchitskaya, G.L., Roy, A., Mohideen U., and
Mostepanenko, V.M. (1999)
Complete roughness and conductivity corrections for the recent
Casimir force measurement,
{\it Phys. Rev.} {\bf A60}, 3487--3497.
\item
Mostepanenko, V.M. and Novello, M. (2001)
Weak scale compactification and constraints on non-Newtonian
gravity in submillimeter range, in A.A.~Bytsenko, 
A.E.~Gon\c{c}alves, and B.M.~Pimentel (eds.),
 {\it Geometric Aspects of Quantum Fields},
World Scientific, Singapore, pp.128--138.
\item
Decca, R.S., L\'{o}pez, D., Fischbach, E., and Krause, D.E.
(2003) Measurement of the Casimir force between dissimilar
metals, {\it Phys. Rev. Lett.} {\bf 91}, 050402-1--4.
\item
Decca, R.S., Fischbach, E., Klimchitskaya, G.L., Krause, D.E.,
L\'{o}pez, D., and Mostepanenko, V.M. (2003) Improved tests of 
extra-dimensional physics and thermal quantum field theory from 
new Casimir force measurements, hep-ph/0310157;
 {\it Phys. Rev. D}, {\bf 68}, N11. 
\item
Dimopoulos, S. and Guidice, G.F. (1996)
Macroscopic forces from supersymmetry,
{\it Phys. Lett. } {\bf B379}, 105--114.
\item
De Sabbata, V., Melnikov, V.N., and Pronin, P.T. (1992)
Theoretical approach to treatment of non-Newtonian forces,
{\it Progress Theor. Phys.} {\bf 88}, 623--661.
\end{enumerate}
\end{article} 
\end{document}